\documentclass[jgr]{agutex}%{AGUTeX}
\usepackage{graphicx}
\usepackage{rotating, lineno}

\setkeys{Gin}{draft=false}

\authorrunninghead{OLIVEIRA AND RAEDER}

\titlerunninghead{SYMMETRY OF IP SHOCKS}

\authoraddr{Corresponding author: Denny Oliveira, EOS Space Science
  Center, University of New Hampshire, Durham, NH USA. (dmz224@wildcats.unh.edu)}

\begin{document}

%----------------------------------------------------------------------------------------
%	TITLE
%----------------------------------------------------------------------------------------

\title{Impact angle control of interplanetary shock geoeffectiveness}

%----------------------------------------------------------------------------------------
%	AUTHORS AND AFFILIATIONS
%----------------------------------------------------------------------------------------

% Use \author{\altaffilmark{}} and \altaffiltext{}

% \altaffilmark will produce footnote; matching \altaffiltext will appear at bottom of page.

\authors{D. M. Oliveira and J. Raeder}

\affil{Department of Physics, and EOS Space Science  Center, University of New Hampshire, Durham, NH USA.}

%\altaffiltext{2}{Department of Geography, Ohio State University, Columbus, Ohio, USA.}

%----------------------------------------------------------------------------------------
%	ABSTRACT
%----------------------------------------------------------------------------------------

% Do NOT include any \begin...\end commands within the body of the abstract.

\begin{abstract}

We use OpenGGCM global MHD simulations to study the nightside magnetospheric, magnetotail, and ionospheric responses to interplanetary (IP) fast forward shocks. Three cases are presented in this study: two inclined oblique shocks, hereafter IOS-1 and IOS-2, where the latter has a Mach number twice stronger than the former. Both shocks have
impact angles of 30$^o$ in relation to the Sun-Earth line. Lastly, we choose a frontal perpendicular
shock, FPS, whose shock normal is along the Sun-Earth line, with the same Mach number as IOS-1.
We find that, in the IOS-1 case, due to the
north-south asymmetry, the magnetotail is deflected southward, leading
to a mild compression. The geomagnetic activity observed in the
nightside ionosphere is then weak. On the other hand, in the head-on
case, the FPS compresses the magnetotail from both sides
symmetrically. This compression triggers a substorm allowing a larger amount of stored energy in the
magnetotail to be released to the nightside ionosphere, resulting in stronger geomagnetic activity. By comparing IOS-2 and FPS, we find
that, despite the IOS-2 having a larger Mach number, the FPS leads to a larger geomagnetic response in the nightside ionosphere. As a result, we conclude that 
IP shocks with similar upstream conditions, such as magnetic field,
speed, density, and Mach number, can have different geoeffectiveness, depending on their shock
normal orientation.

\end{abstract}

%---- SOME DEFINITIONS --------------------------------------------------------------------
%\def\RE{R$_\mathrm{E}$}
%\def\thbn{$\theta_{B_n}$}
%\def\thxn{$\theta_{x_n}$}
%\def\phiyn{$\varphi_{y_n}$}
%\def\lam{$\lambda_m$}

%----------------------------------------------------------------------------------------
%	ARTICLE CONTENT
%----------------------------------------------------------------------------------------

% The body of the article must start with a \begin{article} command
% \end{article} must follow the references section, before the figures and tables.

\begin{article}

\section{Introduction}

Interplanetary (IP) shocks are ubiquitous features of the solar wind \citep{Burlaga1971a}.  As they encounter Earth they interact with the magnetosphere, causing disturbances that can be seen everywhere in the magnetosphere.  Because the disturbances alter the magnetospheric current systems, the ionosphere is also affected, and the magnetic field on the ground is perturbed as well.  The most dramatic shock-induced ground perturbations are the storm sudden commencements (SSCs), which are driven by a strong IP shock preceding a geomagnetic storm driven by coronal mass ejections (CMEs).\par

The interaction of IP shocks with the Earth's magnetosphere is both complex and important.  For example, the shock-shock interactions such as between an IP shock and the Earth's bow shock may occur in many contexts, for example, in the heliosphere and in astrophysical systems.  Such remote interactions are difficult to observe, but can be readily observed with in-situ measurements in the magnetosphere.  On the other hand, strong IP shock impacts on the magnetosphere have substantial space weather effects, for example, they produce geomagnetically induced currents (GICs), which can impact power grids \citep{Bolduc2002,metatech-report-2010}, and they can energize particles in the inner magnetosphere \citep{Hudson1997,Zong2009}.  \cite{Echer2004a} reported that 22\% of all interplanetary shocks are intensely geoeffective while 35\% are moderately geoeffective. Thus, the study of IP shock impacts is of fundamental and also of practical importance.\par

At 1 AU (astronomical unit) IP shocks are almost exclusively fast shocks.  Slow shocks may exist at closer distances to the Sun, but they are subject to Landau damping \citep{Richter1985}.  IP shocks may further be classified as  propagating away from the Sun, i.e., forward shocks, or as propagating towards the Sun, i.e., so-called  reverse shocks.  Since the solar wind speed is almost always supermagnetosonic, a reverse shock will still propagate away from the Sun in the Earth's frame.  IP shocks may then be further classified by their strength, i.e., their Mach number in the solar wind reference frame, their orientation, and by the orientation of their upstream field with respect to the shock normal.  The compression ratio, i.e., the ratio of downstream to upstream plasma density, is an alternative measure for the shock strength, which is often more convenient to use than the Mach number.  Assuming a $\gamma$ (specific heat ratio) of 5/3 appropriate for a monoatomic gas, the compression ratio must lie between 1 and 4 \citep{Priest}.  IP shocks are typically weak (compared to planetary bow shocks) with a compression ratio between 1.2 and 2 \citep{Berdichevsky2000}. The Rankine-Hugoniot relations provide the jump conditions between the upstream and downstream plasma parameters in the MHD context \citep{Jeffrey,Priest,Parks}.  Thus, given the upstream plasma and field parameters, as well as the shock normal, the downstream parameters can easily be calculated.  The inverse calculation, i.e., determining the shock speed and orientation from measured upstream and downstream values, is generally a much more difficult problem, because the critical parameters that cannot be measured directly, such as the shock normal, depend in a very sensitive manner on the upstream and downstream plasma and field measurements.  Near 1~AU, IP shocks can generally be assumed to be planar structures on the scale size of the Earth's magnetosphere.  For example, \cite{Russell1983} found the assumption of planarity consistent with measurements from four widely spaced solar wind monitors. Also using the assumption of shock planarity, \cite{Russell2000a} estimated the shock normal orientation of a large IP shock with accuracy by comparing results of different IP shock normal determination methods.\par

The shock normal vector determines how the shock propagates through the heliosphere. Normals of most IP shocks generated by CMEs at 1 AU are concentrated near the Sun-Earth line \citep{Richter1985}. However, shocks driven by corotating interaction regions (CIRs), as a result of the slow solar wind compression by a fast stream, have normals inclined in relation to the Sun-Earth line \citep[see, e.g.,][and references therein]{Siscoe1976,Pizzo1991}. For CIR-driven shocks, the normal angles in the azimuthal direction in relation to the solar coordinate system are generally equal or larger than the inclination angle \citep {Siscoe1976,Pizzo1991}. \par

Here, we are primarily concerned with the effect that the orientation of an IP shock has on its geoeffectiveness.  Similar studies have been done in the past, but none have addressed the parameter space that we are concerned with. For example, \cite{Jurac2002} analyzed more than 100 IP shocks in terms of the angle between the shock normal and the upstream interplanetary magnetic field (IMF), also called obliquity.  They found that there is about a 40\% chance for an IP shock to be followed by an intense magnetic storm when the obliquity is 90$^o$. These shocks are more likely to result in an intense storm than shocks whose normal are oblique to the IMF direction. \cite{Takeuchi2002b} showed that the shock orientation plays an important role in controlling the rise time duration of an SSC, such  that the rise time is longer in the case of larger impact angles. The unusual SSC rise time observed, $\sim$30 minutes, was explained by a gradual compression observed in the magnetosphere due to the fact that the shock normal had to be largely inclined to the dusk-ward direction. The interaction of IP shocks inclined in relation to the Sun-Earth line with the bow shock was also investigated by \cite{Grib2006}. They solved the Rankine-Hugoniot conditions numerically for different shock inclinations and one of their most important results was that, for shock normal inclinations between 60$^o$ and -60$^o$, the density changed from dusk to dawn in the bow shock in the case where the discontinuity was a fast forward shock (FFS).  Using ACE and Wind satellite data from 1995 to 2004, \cite{Wang2006} reported that, in a survey of nearly 300 FFSs, 75\% of them were followed by SSCs observed on the ground. They also found that the shock impact angle plays an important role in determining the SSC rise time, as previously suggested by \cite{Takeuchi2002b}. When the shock speed (shock strength) was fixed, the more parallel the shock normal with the x-line, the smaller the rise time. The same occurred when they fixed the shock inclination and changed the shock speed. The faster the shock, the shorter the SSC rise time.\par

Several authors have studied the interaction of IP shocks with the Earth's magnetosphere in the context of numerical MHD simulations. However, almost always the IP shock hit the bow shock at the subsolar point head-on. For instance, \cite{Ridley2006} simulated an extreme IP shock driven by a Carrington-like CME \citep{Manchester2006} that pushed the magnetopause toward the Earth to the limit of their code boundary, which was at 2 R$_\mathrm{E}$. They also observed a secondary shock wave reflected back by the magnetopause that encountered the bow shock that was moving inward. Then the combined motion propagated down the flanks of the magnetosphere. Similar bow shock Earth-ward motion was observed by \cite{Safrankova2007} in their numerical simulation for a much weaker IP shock as well. The interaction of IP shocks with the Earth's magnetosphere can also lead to generation of two ionospheric current systems \citep{Ridley2006,Guo2007,Samsonov2010}. The appearance of  an anomalous region I current, which flowed oppositely to the region I current, followed by a frontal IP shock impact with no IMF $B_z$ was found by \cite{Guo2007}. This anomalous region I current formed at noon, developed and then moved toward the evening side until it vanished. Such a response depends on the strength of the IP shock. In other MHD simulations concerning the magnetosheath, three new discontinuities appeared downstream from the bow shock in addition to the FFS: a forward slow expansion wave, a contact discontinuity, and a reverse slow shock \citep{Koval2006,Samsonov2006b,Samsonov2007}. \cite{Samsonov2010} simulated the interaction of an IP shock with the magnetosphere in an artificial case with a northward IMF. The IP shock normal was aligned with the Sun-Earth line. They observed an intensification of two ionospheric current systems (similar to the preliminary and main impulse currents) that coincided in time with the intensification of two corresponding magnetospheric dynamos.\par

\cite{Guo2005} performed a global numerical MHD simulation with different shock normal orientations to study the interaction of IP shocks with the Earth's magnetosphere. They simulated two different cases with a Parker-spiral IMF orientation with no $B_Z$ component.   In their first case, the shock normal was parallel to the Sun-Earth line, and in their second case the shock normal was oblique to this line with an angle of 60$^o$.  They found that the inclined shock normal led to a longer evolution time of the system.   Although both systems evolved from the same initial conditions in their numerical simulations, as can be seen in their Figure 4, they did not find any significant difference in the final quasi-steady state of the systems. Similar results were also found by \cite{Wang2005}. More recently, similar results have been found by \cite{Samsonov2011} as well.  He presented a solution and analysis of the problem of an inclined IP shock incident on and propagating through the Earth's magnetosheath. He showed that inclined IP shocks with normals in the equatorial plane result in a dawn-dusk asymmetry in the magnetosheath and predicted that this effect should be present in observations of sudden impulse inside the magnetosphere and on the ground.  \par

Here, we show that the IP shock normal orientation is a critical parameter determining the geoeffectiveness of IP shocks. Preliminary results of this work were presented by \cite{Oliveira2013}. IP shocks that have their normal aligned with the Sun-Earth line have the largest geoeffectiveness, because they compress the magnetosphere from all sides at the same time.  By contrast, if the shock normal makes a large angle with the Sun-Earth line in the x-z plane in GSE (Geocentric Solar Ecliptic) coordinate system, the north-south asymmetry of the impact pushes the plasma sheet either to the north or to the south without much compression, which leads to a much weaker response.

In the following, we describe the model in section 2, in section 3 we present our results, and in section 4 we summarize and discuss our results.

%----------------------------------------------------------------------------------------
\section{OpenGGCM Model}

We use the Open Geospace General Circulation Model (OpenGGCM) to study the impact of IP shocks on the magnetosphere.  The OpenGGCM is a global coupled model of the Earth's magnetosphere, ionosphere, and thermosphere.  It is available at the Community Coordinated Modeling Center (ccmc.nasa.gov) as a community model for model runs on demand \citep{Rastaetter2013,Pulkkinen2011,Pulkkinen2013}. 
The magnetosphere part solves the MHD equations as an initial-boundary-value problem.
The MHD equations are solved to within $\sim$3 R$_\mathrm{E}$~of Earth.  The region within 3 R$_\mathrm{E}$~is treated as a magnetosphere-ionosphere (MI) coupling region \citep{Oliveira2014a} where physical processes that couple the magnetosphere to the ionosphere-thermosphere system are parameterized using simple models and relationships. The ionosphere-thermosphere system is modeled using the NOAA CTIM (Coupled Thermosphere Ionosphere Model \cite[]{Fuller-Rowell1996,Raeder2001a}).  The OpenGGCM has been described with some detail  in the literature \cite[see, e.g.][]{Raeder2001b,Raeder2003,Raeder2008}; we thus refer the reader to these papers.  The OpenGGCM has been used for numerous studies, including studies of substorms \citep{Raeder2001b,Ge2011,Gilson2012,Raeder2013}, storms 
\citep{Raeder2001a,Raeder2005,Rastaetter2013,Pulkkinen2013},  and, most relevant for this study, for the study of IP shock impacts \citep{Shi2013}.

%----------------------------------------------------------------------------------------
\section{Shock Impacts}

\subsection{Simulation setup}

We use exclusively GSE coordinates in our simulation input data. The numerical box extends from the Earth 30 R$_\mathrm{E}$~in the Sun-ward direction, and 300 R$_\mathrm{E}$ ~down the tail. In the directions perpendicular to the Sun-Earth line, i.e., in the y and z directions, the numerical box extends to $\pm$50 R$_\mathrm{E}$. The numerical grid is non-equidistant Cartesian and is divided into 610$\times$256$\times$256 grid cells, such that the highest resolution is closest to Earth. Specifically, the grid resolution is 0.15 R$_\mathrm{E}$ ~within a distance of 10 R$_\mathrm{E}$ ~radially from the Earth. The inner boundary, where the magnetosphere variables connect via field-line mapping to the ionosphere is located at 3 R$_\mathrm{E}$. \par

In this paper we only consider IP shocks for which the shock normal lies in the GSE x-z plane.  Furthermore, we assume that the IMF also has no GSE y-component.  Thus, the shock geometry relative to the Earth and to the IMF depends exclusively on two angles. First, depending on the shock normal relative to the upstream (relative to the shock) magnetic field direction, a shock can be classified as perpendicular, oblique, or parallel \citep{Burlaga1971a,Tsurutani2011}.  As is often found in the literature \citep{Burlaga1971a,Tsurutani2011}, when $0^o<\theta_{B_n}\leq 30^o$, the shock is classified as almost parallel. In the cases in which $30^o\leq\theta_{B_n}\leq 60^o$, the shock is said to be oblique. Finally, when $60^o\leq\theta_{B_n}<90^o$, the shock is classified as almost perpendicular. In particular for this paper, the shock is named perpendicular when $\theta_{B_n}$ = 90$^o$. Second, in relation to the Earth's system of reference, the shock normal is decomposed in terms of two angles: the angle $\theta_{x_n}$ between the shock normal and the Sun-Earth line, and the angle $\varphi_{y_n}$ in the x-y plane that completes the set, following the notation of \cite{Vinas1986}.  For our simulations presented here, $\varphi_{y_n}$ is always 90$^o$.  We also assume average solar wind conditions, with a particle number density of 5 cm$^{-3}$, thermal plasma pressure of 20 pPa, magnetic field magnitude of 7 nT, and background speed of $v_1$ = (-400,0,0) km/s in all cases upstream of the IP shocks. We specify the shock strength by its compression ratio.  As reported by \cite{Echer2003a}, most shocks near Earth have a compression ratio of the order of 2 during solar maximum. Here, we then choose a compression ratio value of 1.5 in order to have a mild shock. The MHD code input is set as follows. We transform the upstream initial conditions into the shock frame, calculate the downstream parameters, and subsequently transform them back to the Earth's system of reference. We ran simulations of different FFSs with different shock normal orientations, obliquities, shock speeds, Mach numbers, and IMF $B_Z$ pointing either northward or southward. For brevity, we select three FFSs with different shock normal orientations and Mach numbers to discuss in detail. In the first case, the IP shock has its normal inclined with an angle $\theta_{x_n}$ of 30$^o$ with the GSE x-axis toward the south, $\theta_{B_n}$ =51$^o$, shock speed $v_s$ = 380 km/s, and Mach number of 3.7, i.e., an inclined oblique shock, hereafter IOS-1. The second case corresponds to an FFS, also inclined and oblique here called IOS-2, with $\theta_{x_n}$ = 30$^o$, $\theta_{B_n}$ = 45$^o$, $v_s$ = 650 km/s, and Mach number of 7.4. In the last case the shock normal has $\theta_{x_n}$ = 0$^o$ and was perpendicular to the IMF, i.e.,  a frontal perpendicular shock (FPS), with $v_s$ = 650 km/s, and Mach number of 3.7. These shock speeds are consistent with the observations reported by \cite{Berdichevsky2000}, where most FFSs have speeds in the range 50-200 km/s in the shock frame of reference. Tables 1,2, and 3 show the upstream (1) and downstream (2) and other important parameters for the three FFSs. The first and second shocks impact the magnetosphere (first contact with the magnetopause) at t = 16.45 minutes, and the FPS reaches the subsolar magnetopause at t=18.28 minutes. We also simulated shocks with northward IMF and otherwise identical solar wind conditions.  We found that the results were similar to the southward cases, but with weaker magnetosphere response,  with the exception that transient northward B$_z$ (NBZ) currents occurred within a few seconds after the shock impact when it was frontal.  This effect was already reported by \cite{Samsonov2010}.  We will thus focus on the case of southward IMF.

%----------------------- Table 1
\begin{table}
\begin{tabular}{l c c c c c c}
\hline
\hline
\multicolumn{7}{c}{{\bf IOS-1}, $v_s$ = 380 km/s, M = 3.7} \\
\multicolumn{7}{c}{$\theta_{B_n}$ = 51$^o$, $\theta_{x_n}$ = 30$^o$} \\
\hline
 & $B_{x}$ & $B_{z}$ & $v_{x}$ &  $v_{z}$ & $P$ & $n$ \\
\hline
 upstream & -1.83 & -6.83 & -400.00 & 0.00 & 20.0 & 5.0 \\
downstream & -0.52 & -9.09 & -434.15 & -17.65 & 67.45 & 7.5\\
\hline
\hline
\end{tabular}
\caption{Upstream and downstream plasma parameters for the inclined oblique case IOS-1 with shock speed of 380 km/s and impact angle $\theta_{x_n}$ = 30$^o$.  Upstream Mach number and obliquity are also shown.\tablenotemark{a}}\vspace{0.0cm}
\tablenotetext{a}{B in nT, v in km/s, P in pPa, and n in particles/cm$^3$.}
\end{table}

%----------------------- Table 2
\begin{table}
\begin{tabular}{l c c c c c c}
\hline
\hline
\multicolumn{7}{c}{{\bf IOS-2}, $v_s$ = 650 km/s, M = 7.4} \\
\multicolumn{7}{c}{$\theta_{B_n}$ = 45$^o$, $\theta_{x_n}$ = 30$^o$} \\
\hline
 & $B_{x}$ & $B_{z}$ & $v_{x}$ &  $v_{z}$ & $P$ & $n$ \\
\hline
 upstream & -1.83 & -6.83 & -400.00 & 0.00 & 20.0 & 5.0 \\
downstream & -0.52 & -9.09 & -461.53 & -28.61 & 109.74 & 7.5\\
\hline
\hline
\end{tabular}
\caption{Same plasma parameters for the inclined oblique case IOS-2 with shock speed of 650 km/s and the same impact angle. }\vspace{0.0cm}
\end{table}

%----------------------- Table 3
\begin{table}
\begin{tabular}{l c c c c c c}
\hline
\hline
\multicolumn{7}{c}{{\bf FPS}, $v_s$ = 650 km/s, M = 3.7} \\
\multicolumn{7}{c}{$\theta_{B_n}$ = 90$^o$, $\theta_{x_n}$ = 0$^o$} \\
\hline
 & $B_{x}$ & $B_{z}$ & $v_{x}$ &  $v_{z}$ & $P$ & $n$ \\
\hline
 upstream & 0.00 & -7.07 & -400.00 & 0.00 & 20.0 & 5.0 \\
downstream & 0.00 & -10.61 & -483.33 & 0.00 & 191.37 & 7.5\\
\hline
\hline
\end{tabular}
\caption{Same plasma parameters for the frontal perpendicular case FPS with shock speed of 650 km/s and impact angle $\theta_{x_n}$ = 0$^o$. }\vspace{0.0cm}
\end{table}

%----------------------------------------------------------------------------------------
\subsection{Results}

As the IP shock impacts the magnetopause, it launches waves into the magnetosphere.  The phase speed of these waves (both Alfv\'en and magnetosonic waves) is generally much larger in the magnetosphere than in the solar wind and the magnetosheath.  Therefore, the amplitude of the waves diminishes in the magnetosphere, because the waves are partially reflected and also because of the higher phase speed the wave energy spreads out more quickly.  In order to visualize such waves, we therefore subtract consecutive time instances from each other to remove the background as much as possible.  This essentially amounts to taking the time derivative.  Thus, in the plots shown below, for any quantity $X$, we show the difference $\Delta X = X(t) - X(t-\Delta t)$, where $\Delta t$ is chosen to be 30 seconds. 
Figure~1 shows the time evolution of the total magnetic field changes ($\Delta$B), in nT, in the noon-midnight meridian plane. The left column shows the IOS-1 case, the middle column shows the IOS-2 case, and the right column shows the FPS case. Each row shows different time, and the time difference between rows is three minutes. The red color in the color bar shows an increasing magnetic field $B$, and the blue color indicates where $B$ decreases. The color bar range is $\pm$2 nT. In all columns, the first plots show the instant right after the FFSs crosses the bow shock. 
%As the waves move through the magnetosheath, they propagate around the flanks of the magnetosheath.

%\vspace{1cm}
\begin{figure*}[h]
\vspace{-0.8cm}
\hspace*{1.0cm}\includegraphics[width=0.87\hsize]{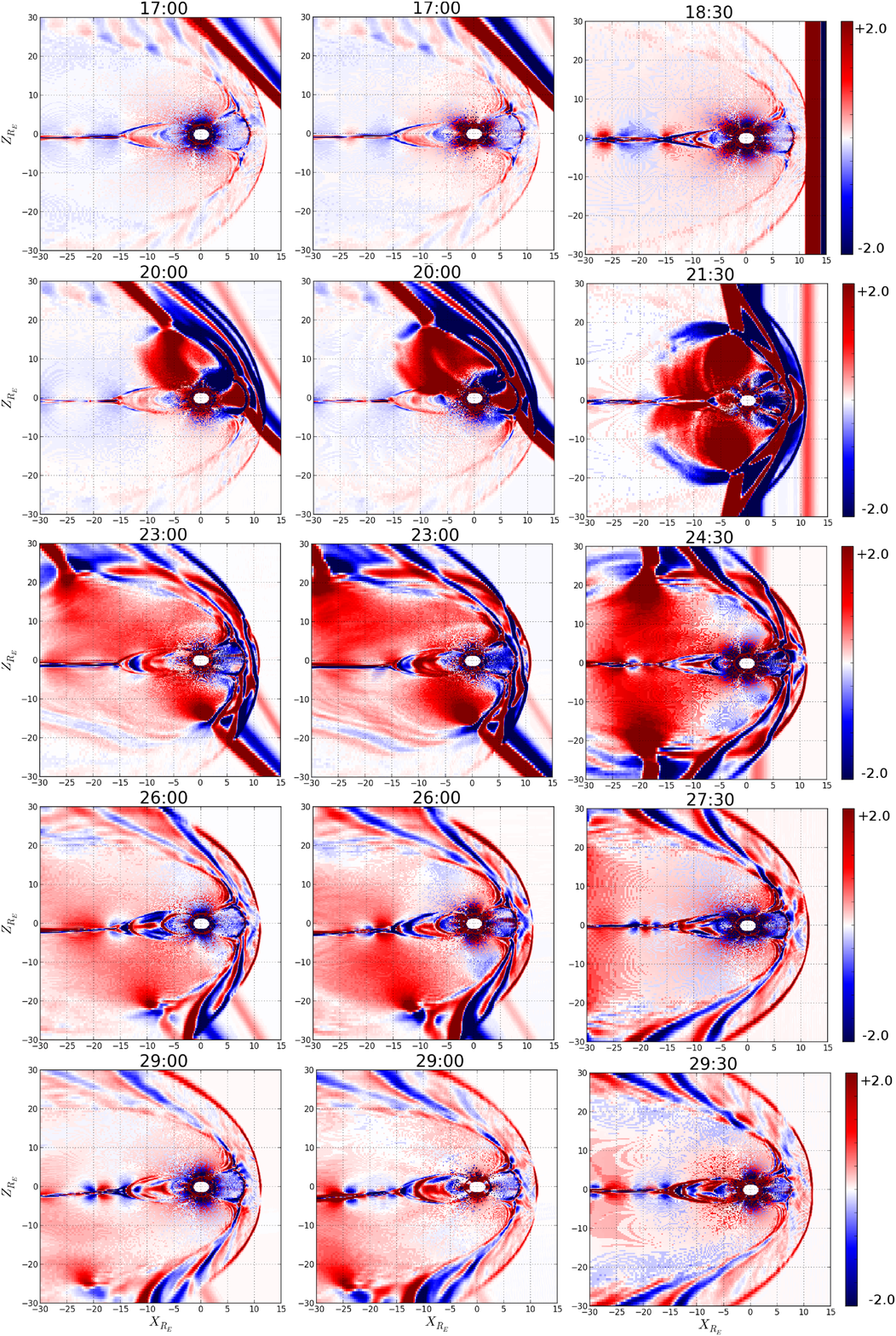}
\vspace{0.125cm}
\caption{5 consecutive frames representing the result of numerical simulations of $\Delta$B(nT) plotted in the meridian plane with $\Delta$t = 3 minutes, in a time range of 12 minutes. From left, first column shows the IOS-1 case, central column represents the IOS-2 case, and the last column shows the FPS case. See text for details.}
%\label{figure_label}

\vspace{-0.8cm}

\end{figure*}
\begin{figure*}[h]
\vspace{-0.8cm}
\hspace*{2.11cm}\includegraphics[width=0.72\hsize]{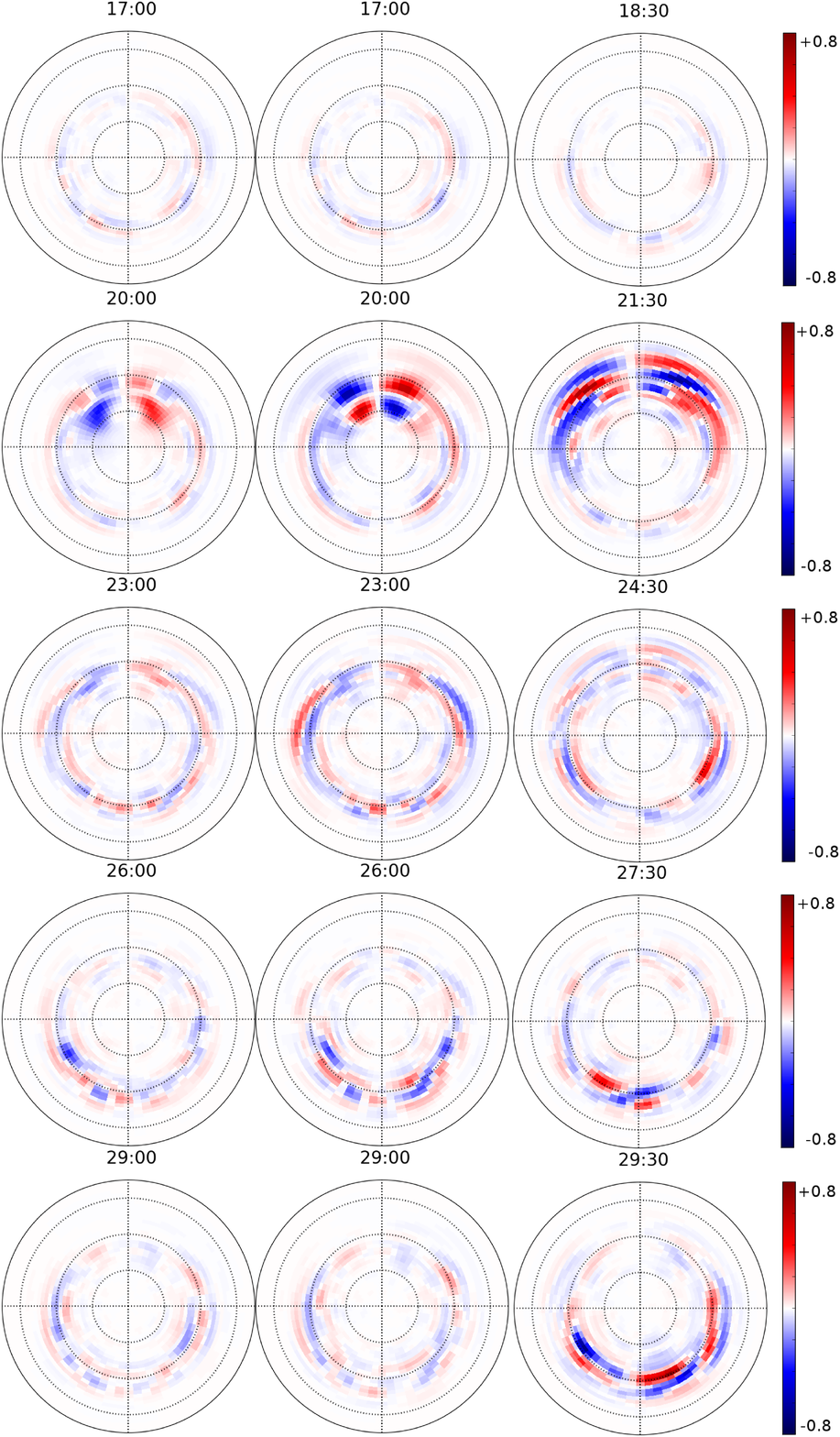}
\vspace{-0.105cm}
\caption{Difference of field-aligned currents, $\Delta$FAC($\mu$A/m$^2$), for the northern hemisphere ionosphere in the same sequence as represented in Figure 1. The center of each plot is the magnetic pole. The right side of each plot is dawn while the top is noon (or towards the Sun), and the bottom is the midnight. See text for details. }
%\label{figure_label}
\end{figure*}
\begin{figure*}[h]
\vspace{-1.08cm}
\hspace*{2.1cm}\includegraphics[width=0.81\hsize]{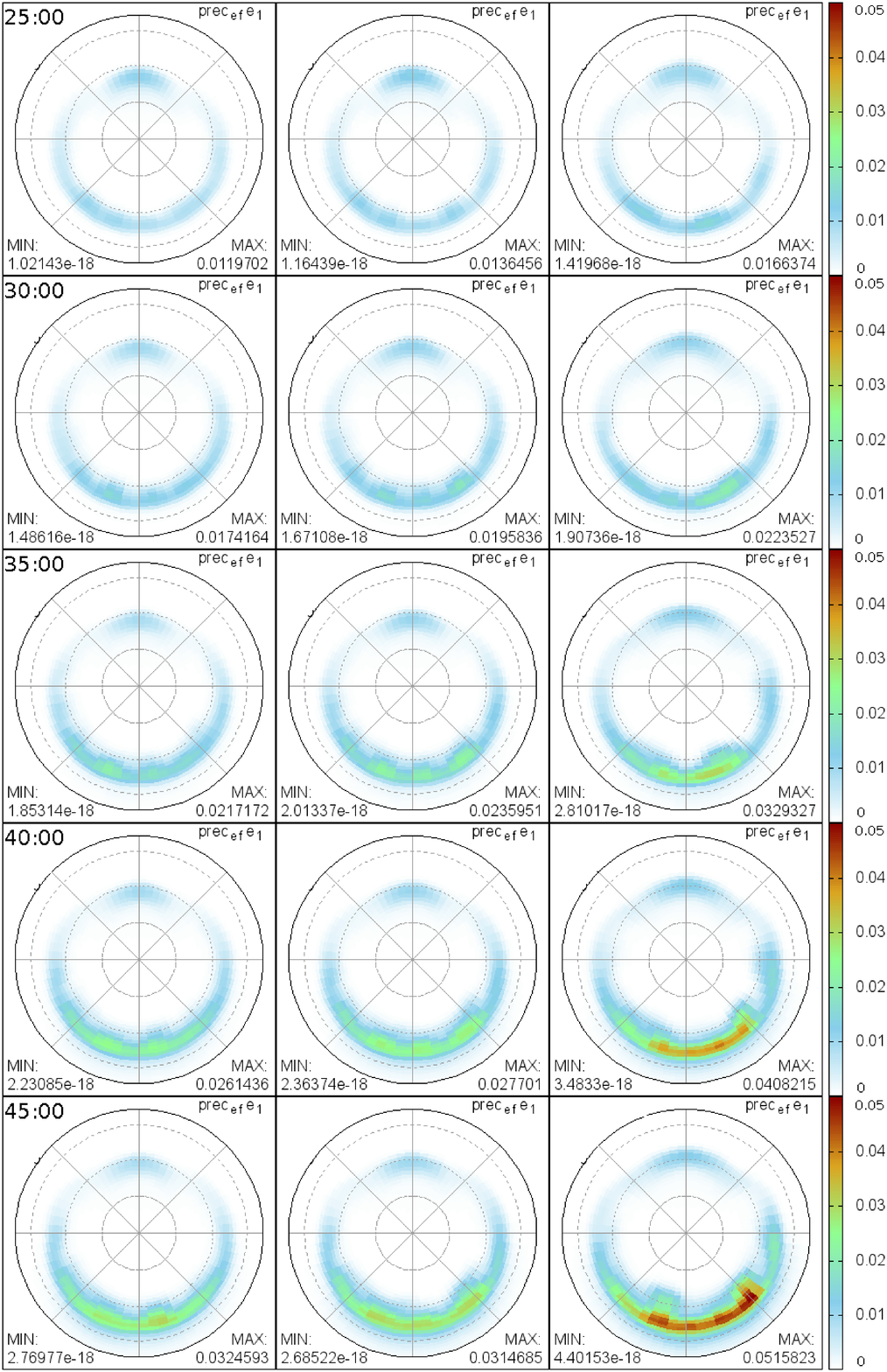}
\vspace{-0.11cm}
\caption{DAPEF(mW/m$^2$) plotted in a time range of 20 minutes with time interval of 5 minutes. The first column represents the IOS-1 case, second the IOS-2 case, and third the FPS case. Note that only the FPS case shows the occurrence of a substorm onset. See text for more details.}
%\label{figure_label}
\end{figure*}

\vspace*{-0.38cm}

In Figure~1 the shock fronts appear much broader than they really are.  First, by taking differences 30 seconds apart, the shock propagates ~2-3 R$_\mathrm{E}$~ across the grid during that time, thus the shock will appear at least that wide.  Second, because of numerical diffusion, the shocks have a foot or ramp on either side of the shock front.  Normally, this would be hardly visible in a color-coded plot.  However, because we take the differences and clamp the color bar at low values, these numerical artifacts become emphasized and make the shock appear much wider than it really is.  In the simulation, the shocks are resolved to within 2-3 cells, i.e., to less than 0.5 R$_\mathrm{E}$~within 10 R$_\mathrm{E}$~from the x-axis.

As soon as the shock impacts the magnetopause, it launches Alfv\'en waves and magnetosonic waves into the magnetosphere.  Because the wave speeds are much higher in the magnetosphere, these waves race ahead of the IP shock in the magnetosheath.  The second row shows for both cases the time just after the IP shock has impacted the magnetopause.  In the IOS-1 and IOS-2 cases, the first contact between the shock and the magnetopause occurs just past the northern cusp.  The induced wave propagates through the northern lobe and reaches the plasma sheet in the nightside from the north.  At this time, the IP shock has not yet impacted the southern hemisphere, and consequently there is no corresponding wave in the southern hemisphere yet.

The FPS case (second row, right panel) is distinctively different.  The impacts on the northern and southern lobes occur simultaneously, and symmetric waves are launched from both the northern and southern magnetopause into the magnetosphere.   These waves converge on the tail plasma sheet and cause a more significant compression of the plasma sheet.  

As time progresses, these waves propagate further tail-ward.  In the asymmetric cases, the waves from the southern hemisphere reach the near-Earth plasma sheet approximately 3 minutes after the waves from the northern hemisphere.  As a result of such asymmetric impact, there is much less compression of the plasma sheet.  Instead, the entire plasma sheet is bent southward.  This is best visible at the later times (fourth and fifth row), where the deflection at x=-30 R$_\mathrm{E}$~is as much as 3 R$_\mathrm{E}$~in the IOS-1 case and 4 R$_\mathrm{E}$~in the IOS-2 case.  By contrast, in the symmetric case there is no such deflection, but instead a transient compression of the plasma sheet.

After the IP shock has passed (bottom row), the magnetosphere state seems to be very similar for the three cases.  However, the particular display only shows differences to highlight transients and thus provides little information abut the state. \par

In order to examine the effects of the IP shocks on geomagnetic activity, we examine relevant ionospheric quantities.
Figure~2 shows the time differences of the field-aligned current density (FAC, in $\mu$A/m$^2$) in the northern hemisphere polar cap, displayed in the same way as the magnetic field evolution of Figure~1. The times in Figure~2 correspond to the times shown in Figure~1, i.e., a time range of 12 minutes, plotted in three minute increments. The range in the color bar is $\pm$0.8 $\mu$A/m$^2$, and regions in red indicate a positive change in FAC, while regions in blue indicate a negative change in FACs.  The dashed circles represent the magnetic latitude $\lambda_m$ in 10 degree increments from  55$^o$ to the pole.  Left, middle, and right columns represent the results for the IOS-1, IOS-2, and FPS cases, respectively. In all cases, in the first plot, the ionosphere is steady because the FFSs have not yet impacted the magnetopause. Also, in the three situations, the first FAC changes are seen about three minutes after the shock impact, as can be seen at t = 20:00 minutes for the inclined cases and t = 21:30 minutes for the FPS case.  In the two IOS cases, the FAC changes are mostly in the vicinity of the cusp, whereas the FFS causes a broader signature that encompasses the entire dayside auroral region. The changes in the IOS-1 and IOS-2 cases are very similar geometrically, with the only difference that the IOS-2 response is slightly stronger than the IOS-1 response.

As time evolves, with the exception of the first minutes after the shock impact, the FAC signature diminishes in all cases, as can be seen in the middle row.  Subsequently, activity in the nightside develops for the three cases.  However, in the FPS case, the ionosphere response is much stronger.  At t = 24:30 minutes, the enhancement of FACs is most evident on the nightside of the ionosphere between $\lambda_m$ = 70$^o$ and $\lambda_m$ = 65$^o$, and between 0300 MLT and 0600 MLT. At t = 27:30 minutes and t = 29:30 minutes, FACs are enhanced close to midnight local time, which is a typical substorm signature \citep{Akasofu1964,McPherron1991}. The last row of Figure~2 shows the strongest nightside  FAC variations occurring for the FPS case at t = 29:30 minutes, covering almost all the nightside ionosphere for $\lambda_m$ between 70$^o$ and 65$^o$. We attribute this activity due to substorm activity that was triggered by the converging waves in the plasma sheet. \par

Figure~3 shows the diffusive auroral e$^-$ precipitation energy flux (DAPEF) in the time steps of 5 minutes from t = 25:00 minutes. In the IOS-1 and IOS-2 cases the DAPEF remained almost unaffected by the shock impact.  In all cases, auroral precipitation is enhanced in the night side ionosphere approximately 12 minutes after shock impact. In the IOS cases, the shocks hit the magnetosphere behind the cusp, leading to a mild auroral precipitation at t = 35:00. However, at the same instant, the FPS enhances more auroral activity in the ionosphere nightside because the FPS impacts the magnetosphere behind the cusp, and thus the cusp is neither displaced nor compressed. In the FPS case, the shock hits the magnetosphere first at the nose, leading to a compression of the magnetosphere and the cusp, and to a pole-ward displacement of the cusp. \par

Later precipitation changes all occur in the nightside.  As already shown by the FACs, these changes are mostly related to substorm activity. In the FPS case, auroral substorm onset is formed in the ionosphere nightside at  70$^o<\lambda_m<65^o$ between 0300 MLT and 0600 MLT. Such auroral activity is not found in any IOS case. \par

%\vspace{-0.3cm}

In order to perform a quantitative comparison we integrate the ionosphere quantities over the northern hemisphere polar cap.  To separate the directly driven response, which occurs mostly in the dayside, from the induced substorm response, which affects mostly the nightside, we integrate the FACs separately for the dayside and the nightside.

Figure~4 shows the time series of the integrated FACs, in MA, over the northern hemisphere ionosphere on both dayside (a) and nightside (b), for the three cases.  In both panels, the blue lines indicate the IOS-1 case, the green lines indicate the IOS-2 case, and the red lines indicate the FPS case. The first vertical dashed line indicates the instance at which the IP shock hits the bow shock in the IOS-1 and IOS-2 cases, and the second vertical dashed line indicates that instance for the FPS case. The plotted quantity is the magnitude of the FAC, which is primarily the Region I current. As expected, the IP impact enhances the FACs on the dayside. Before the shock impact, the FACs on the dayside have nearly the same magnitude, 0.29 MA, and were in a quasi-steady state. In all cases, the dayside FAC magnitude begins to rise approximately 6 minutes after the shock impact.  The FACs increase nearly linearly over a period of approximately 4 minutes, which corresponds roughly to the time it takes for the shock to pass over the dayside magnetosphere.  This initial rise is nearly identical for all cases.  Later, in the two IOS cases, the FAC remains by and large steady at around 0.5 MA for the IOS case, but continues to rise for another 20 minutes in the FPS case.  \par

Figure~2 shows that the initial rise is in all cases due to enhanced FACs in the vicinity of the cusp.  However, in the IOS-1 and IOS-2 cases the enhancement is mostly in the nightside of the cusp, whereas in the FPS case FACs both on the dayside and the nightside of the cusp are enhanced.  It seems that the more thorough compression of the dayside magnetosphere leads to the stronger and more long lasting FAC enhancement in the FPS case, although that is not directly apparent in Figure~2. 

\begin{center}
\begin{figure}
\vspace{-0.3cm}
\hspace*{-0.27cm}\includegraphics[width=1.04\hsize]{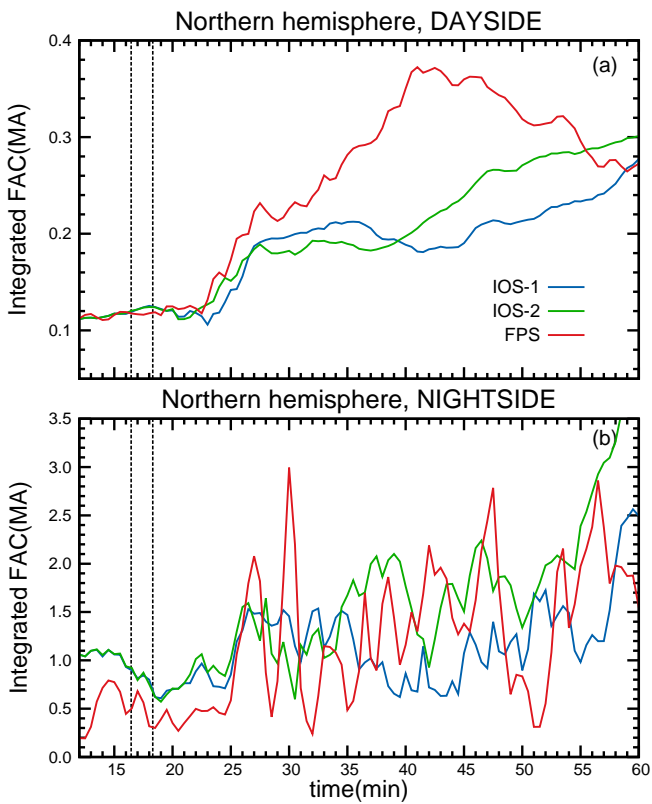}
\vspace{0.05cm}
\caption{Total field-aligned currents (FACs, in MA) integrated in the northern hemisphere ionosphere. The top panel (a) shows the time evolution of the integrated FACs in the dayside ionosphere. The bottom panel (b) represents the integrated FACs in the nightside ionosphere. In both plots, the first dashed vertical line at t=16:45 min indicates the instant of impact of the IOS-1 and IOS-2. The second dashed vertical line, at t=18:28 min, indicates the instant of impact of the FPS. }
%\label{figure_label}
\end{figure}
\end{center}

The nightside current enhancements (Figure~4~(b)) begin about 2 minutes after the dayside enhancements, in the three cases.  For all shocks, the enhancements are qualitatively different from those on the dayside.  In all cases, strong ULF waves are excited, with a period of 4-5 minutes.  In the two IOS cases, the average FAC is enhanced from $\sim$2~MA to $\sim$3~MA due to the shock impact, with superimposed waves with an amplitude of the order of $\sim$0.2-0.3~MA. In the IOS-2 case, 17-18 minutes after the shock impact, the ULF wave amplitude rises to $\sim$0.75~MA.  In the FPS case, the increase of the average FAC is similar, but the wave amplitude is much larger, i.e., of the order of $\sim$1-2~MA.  These large oscillations persist for at least 60 minutes.  While much of the response is similar to what \cite{Guo2005} found in their simulations, the oscillations are new.  It is at present not clear what these oscillations correspond to in the magnetosphere, but they are likely cavity modes \citep{Samson1992}. \par

Figure 5 shows the cross polar cap potential (CPCP) and the integrated precipitation energy DAPEF in panels (a) and (b), respectively. DAPEF is calculated as the thermal energy flux of plasma sheet electrons assuming perfect pitch angle scattering, and a full loss cone \citep[see][for details]{Raeder1998,Raeder2003}.The blue line represents the IOS-1 case, the green line represents the IOS-2 case, and the red line represents the FPS case. Vertical, dashed lines, as described above,  indicate the shock impact times for all cases. 
Before the shock impact, the system oscillated noticeably in an amplitude less than 5 kV in both cases. After the impact, the IOS-1 and IOS-2 induced a very similar potential jump from roughly 35~kV to a peak of approximately~55-60 kV.  The CPCP oscillated with a period of nearly 5 minutes until it reached the near steady state value of 25 kV after 42 minutes. 

The FPS case is more dynamic. The system oscillated around 30~kV before the shock impact. Right two minutes after the FPS hit the bow shock, the potential difference dropped to 23 kV. This effect was also seen by \cite{Guo2005} and interpreted as the redistribution of the FACs after the shock penetrates the magnetosphere. The potential drop then reached two smaller peaks, 50~kV at t=22:00 minutes, and 62~kV at t=24:00 minutes until it reached the maximum of 82~kV at t=27:00 minutes. The potential still reached two peaks of 53~kV at t=30:00 minutes and t=33:00 minutes. Then, the potential difference decreased in a period of nearly 7 minutes. After close to t=33:00 minutes, both systems evolved to nearly the same final quasi-steady state. This similarity has been shown by \cite{Guo2005} but not with this oscillatory behavior. \par

The bottom panel (b) of Figure~5 shows the time series for the integrated DAPEF, in units of GW. Again, the blue line represents the IOS-1 case, the green line represents the IOS-2 case, and the red line represents the FPS case. Before the shock impact the ionosphere is in a quiet state with low auroral activity. After the shock hit the bow shock, between t=24:00 minutes and t=33:00 minutes, the three systems evolved quite similarly, while the auroral energy flux suffered a small drop in the FPS case at t=34:00 minutes. In the IOS cases, the DAPEF attains a maximum value of barely 100-120 GW near t=44:00 minutes, and then evolves to a final state of $\sim$20-50 GW. On the other hand, the FPS enhanced the DAPEF much more, and briefly reaches a maximum of $\sim$350~GW at t=45:00 minutes. \par

By comparing the two shocks with the same Mach number, namely IOS-1 and FPS, we observe that both shocks lead to very different geomagnetic responses. Also, the IOS-2 geomagnetic response is smaller in comparison to the FPS geomagnetic response, even though the inclined shock was twice stronger than the head-on shock. We attribute these results to the different shock normal orientations. Comparison with Figure~3 shows that this enhanced precipitation flux must be from the nightside due to substorm activity.
Such precipitation enhancements have been reported earlier by \cite{Zhou2001}, \cite{Tsurutani2003a}, and \cite{Yue2010}, who find that the auroral precipitation and field-aligned currents in the nightside can be intensified after a FFS impact, because it triggers the release of stored magnetospheric/magnetotail energy in the form of particularly large substorms, or even supersubstorms \citep[See][for more references therein]{Tsurutani2014a}. Such substorm triggering might be a result of the decreasing in the nightside $B_Z$, as observed at geosynchronous orbit by \cite{Sun2011,Sun2012}. In that case, the decrease in the nightside $B_Z$ is suggested as a result of Earth-ward transportation of magnetic flux by temporarily enhanced plasma flows at the nightside geosynchronous orbit after the impingement of an IP shock. In our case, the symmetric plasma transport leads to a more intense geomagnetic activity. Although our simulations do not represent a storm and although the simulation results may be qualitatively in error, they match qualitatively the observed  shock impact behavior. 
\par

\begin{figure}
\vspace{-0.15cm}
\hspace*{-0.27cm}\includegraphics[width=1.04\hsize]{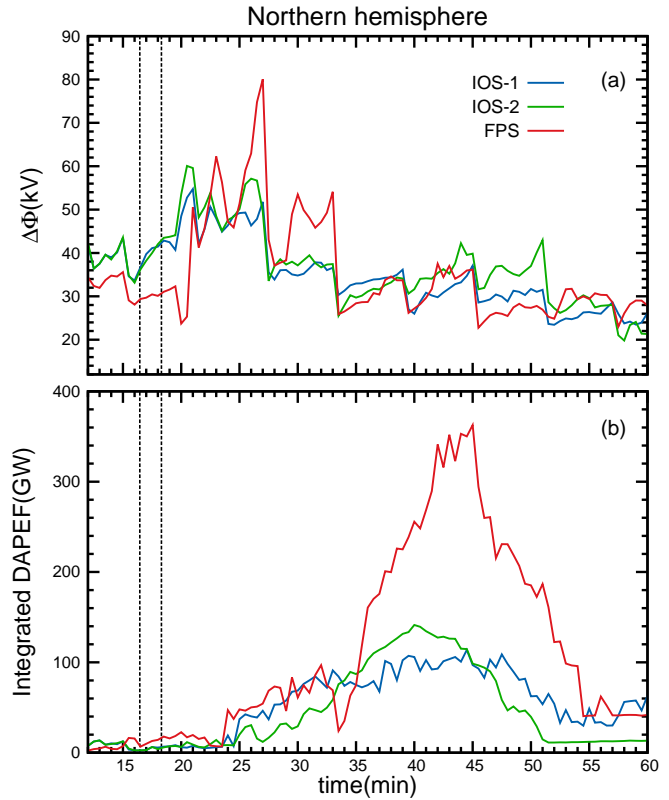}
\vspace{0.0cm}
\caption{(a) Cross polar cap potential (CPCP, in kV) in the northern hemisphere ionosphere.  (b) Integrated auroral electron precipitation energy flux (DAPEF, in GW) in the same hemisphere. The dashed vertical lines are the same as in Figure 4.}
%\label{figure_label}
\end{figure}

%----------------------------------------------------------------------------------------
\section{Summary and Conclusions}

It has been known for a long time that IP shocks can have a profound impact on the magnetosphere; however, it is much less known which factors determine the geoeffectiveness of IP shocks.  There have been a few studies in the past addressing the effect of shock geometry \citep{Takeuchi2002b,Jurac2002,Guo2005,Wang2005,Wang2006,Grib2006,Samsonov2011}, but none have considered the particular geometry that we investigated here. Specifically, here we use global simulations to contrast two different scenarios.  In one scenario, the IP shock normal lies along the Sun-Earth line, such that there is a frontal impact on the magnetosphere.  In the other scenario, the IP shock is inclined with respect to the Sun-Earth line.  In either case, the shock normal lies in the GSE x-z plane, and the IMF is southward, such that there is no y-dependence of any solar wind parameter.  The two scenarios lead to very different responses of the magnetosphere:

\begin{enumerate}

\item  In the frontal case, the shock launches waves symmetrically into the magnetosphere, which converge on the tail plasma sheet and compress it.  By contrast, in the inclined cases, the waves reach the plasma sheet at different times, causing much less compression, but a north-south displacement of the plasma sheet. This result holds even for shocks with larger Mach numbers.

\item  In the frontal case, the compression triggers a substorm, whereas in the inclined cases there is no excess geomagnetic activity beyond what would be expected for southward IMF.

\item  In all cases the shock impact enhances FACs in the dayside with a similar quantitative response.  However, in the inclined cases the FAC enhancement occurs mainly behind the cusp, whereas in the frontal case the cusp is displaced while the FACs increase over a wider MLT range.  In the frontal case, the dayside FAC response also persists longer and is more intense, i.e., a $\sim$200\% enhancement versus a $\sim$100\% enhancement in the inclined case.

\item   The nightside FAC response is qualitatively similar in the three cases and shows the development of ULF waves.  However, in the frontal case the ULF wave amplitude is much stronger.  The detailed excitation mechanism remains to be investigated.

\item  The response of the cross polar cap potential is relatively benign and limited to the first 15 minutes after the impact, in all cases.  The three cases relax to the pre-impact state in less than 20 minutes.

\item  In all cases diffuse auroral electron precipitation increases in similar fashion in direct response to the shock impact.  In the frontal case, this is followed by a delayed response $\sim$10~minutes later, which peaks $\sim$20~minutes after the impact, and which comes from the nightside.  The latter is interpreted as a consequence of the substorm triggered by the shock impact.
\end{enumerate}  

Our results show that the shock impact angle has a major effect on the geoeffectiveness of the shock, even more  than the Mach number or some other measure of the shock strength.  Although we only covered a relatively small parameter space in terms of impact angles, shock strength, and IMF orientation, the qualitative and quantitative differences we found are significant.  With respect to substorm triggering \citep{Kokubun1977,Lyons1995,Lyons1996,Lui1990}, the differences we found in the inclined and frontal cases are of particular importance.  Apparently, the same type of IP shock can either trigger a substorm or not, depending on the shock normal direction, which has not been considered in previous studies.  We also find large amplitude waves in the FACs that are apparently caused by the shock impacts.  Given their period, these waves are likely cavity modes \citep{Samson1992}.  We find that the modes have significantly larger amplitude for the frontal case.  This is likely due to the fact that the waves that converge on the tail and compress the plasma sheet from there launch a wave back towards the nightside magnetosphere, which in turn excites the cavity mode.  In the inclined cases, this earthward wave is likely much weaker, and thus excites a weaker cavity mode.  We will study the wave excitation in more detail in a forthcoming paper.  Other future work is also clearly laid out, namely finding the correlation between geomagnetic activity and IP shock normal orientation in data, and a better parameter space coverage with simulations.

\begin{acknowledgments}
This work was supported by grant AGS-1143895 from the National Science Foundation and grant FA-9550-120264 from the Air Force Office of Sponsored Research. Computations were performed on Trillian, a Cray XE6m-200 supercomputer at UNH supported by the NSF MRI program under grant PHY-1229408. The artificial solar wind data input files from the simulations can be obtained from the corresponding author at dennymauricio@gmail.com.
\end{acknowledgments}

%\bibliography{DO}{}
%\bibliographystyle{agufull08}

\end{article}

\end{document}